\documentclass[twocolumn]{aastex701}
\usepackage[utf8]{inputenc}
\usepackage{natbib}
\usepackage{graphicx}
\usepackage{xcolor}
\usepackage{threeparttable}
\usepackage{multirow}
\usepackage{lipsum}

\received{}
\revised{}
\accepted{}
\submitjournal{ApJL}

\shorttitle{Limits on LHS 1140\,{\rm b} Helium with SOSS}
\shortauthors{Radica et al.}

\begin{document}

\title{Strict Limits on Helium Absorption from LHS 1140\,b from Four JWST NIRISS Transits}

%========== Author List ==========
\author[0000-0002-3328-1203]{Michael Radica}
\altaffiliation{NSERC Postdoctoral Fellow}
\affiliation{Department of Astronomy \& Astrophysics, University of Chicago, 5640 South Ellis Avenue, Chicago, IL 60637, USA}
\email[show]{radicamc@uchicago.edu}

%========== Abstract ==========
\begin{abstract}
Orbiting in the habitable zone of its host star, the 1.7\,$R_\oplus$, 5.6\,$M_\oplus$ planet LHS 1140\,b is a target of great interest. Recently \citet{cherubim_helium_2026} published a detection of metastable He escaping from the atmosphere of LHS 1140\,b, simultaneously providing the first inference of an atmosphere on this planet and indicating that the atmosphere is He-rich and H-poor as would be expected due to Gyrs of fractionated mass loss. In this work, we analyze four archival transits of LHS 1140\,b, spanning Dec 2023 to Jul 2026, taken with the NIRISS instrument on JWST, for evidence of He absorption. Each of the four visits disfavours the presence of He absorption compared to a flat continuum with odds ratios ranging from 3.9--11.6:1. He absorption with an amplitude and width equivalent to that observed by \citet{cherubim_helium_2026} is strongly ruled out by the data with odds ratios from 300--8.6$\times$10$^4$:1 compared to a flat continuum --- though it should be noted that none of the JWST transits are contemporaneous with the \citet{cherubim_helium_2026} detection. We also fit the absolute out-of-transit stellar spectra from these four visits, as well as an additional JWST NIRISS transit of planet c, to search for evidence of stellar variability. The stellar properties inferred from the first visit are discrepant with the other four, potentially indicating evidence for variability. Further monitoring is necessary to confirm this. In all, our work provides a set of strict limits on He escape from LHS 1140\,b that will be valuable to future studies into the nature and evolution of this intriguing world.
\end{abstract}


% ======== Main Text =========
% === Introduction ====
\section{Introduction} 
\label{sec: Introduction}

The LHS 1140 system consists of two small planets orbiting a quiet, mid-M dwarf star. The outer planet, LHS 1140\,b has garnered significant interest from the exoplanet community due to the fact that it orbits in the habitable zone of the host star \citep{dittmann_temperate_2017, ment_second_2019, lill-box_planetary_2020, cadieux_new_2024}. Originally thought to be a terrestrial, potentially Earth-like planet, the revisement of its mass (5.6\,$M_\oplus$) and radius (1.7\,$R_\oplus$) by \citet{cadieux_new_2024} make LHS 1140\,b more likely to be a temperate mini-Neptune (i.e., with a $\sim$0.1\% H/He envelope) or a water world --- the latter of which retains a potential for habitability \citep[e.g.,][]{kite_habitability_2018, cadieux_new_2024, damiano_lhs1140_2024}. 

LHS 1140, the star, is known to be an old and relatively inactive M dwarf \citep{dittmann_temperate_2017, cadieux_new_2024} with a low flare rate \citep{medina_galactic_2022}. This fact, combined with LHS 1140\,b's relatively large escape velocity place LHS 1140\,b strongly on the atmosphere favoured side of the cosmic shoreline --- the theoretical line in stellar XUV flux-escape velocity space that divides planets that can retain atmospheres from those that cannot \citep{zahnle_cosmic_2017, pass_receding_2025, meni-gallardo_empirical_2025}. 

For this reason, LHS 1140\,b has been the target of several atmosphere searches using both space- \citep[e.g.,][]{edwards_hubble_2021, damiano_lhs1140_2024, cadieux_transmission_2024} and ground-based \citep[e.g.,][]{diamond-lowe_simultaneous_2020} instruments. JWST observations disfavour low-mean-molecular-weight (MMW) H/He-dominated atmospheres and provide tentative hints of high-MMW envelopes \citep{damiano_lhs1140_2024, cadieux_new_2024}, though these inferences are complicated by the effects of unocculted stellar surface heterogeneities (i.e., the transit light source (TLS) effect; \citealp{rackham_transit_2018, rackham_transit_2019}).

Recently, \citet{cherubim_helium_2026} presented two transits of LHS 1140\,b taken with the high-resolution ($R$$\sim$68,000) 
Warm Infrared Echelle Spectrograph to Realize Extreme Dispersion (WINERED) on the Magellan Clay Telescope. The first transit, taken in Sept 2024, displayed a clear signature at 1.083\,µm attributed to the metastable He triplet --- a strong tracer of atmosphere escape \citep[e.g.,][]{spake_helium_2018, oklopcic_new_2018, oklopcic_helium_2019}. The two strongest lines of the triplet (which are blended in the WINERED observations) are well fit by a Gaussian line profile with an amplitude of $1.24^{+0.22}_{-0.23}$\% and a full-width-half-max (FWHM) of $0.86^{+0.15}_{-0.27}$\,\AA. Hydrodynamic escape modelling yielded an estimated mass-loss rate of 3$\times$10$^8$\,g/s and a H/He ratio of $\sim$1$\times$10$^{-3}$. This simultaneously provided the first detection of an atmosphere on this planet and indicated that it has a He-rich composition, potentially consistent with the previous JWST transmission observations \citep{cadieux_transmission_2024, damiano_lhs1140_2024} and expectations of fractionated mass loss \citep[e.g.,][]{cherubim_oxidation_2025}.

However, the second WINERED transit in Sept 2025 yielded a non-detection of escaping He, with \citet{cherubim_helium_2026} reporting an upper limit on the observable He line depth of 0.6\%. \citet{cherubim_helium_2026} attribute this inconsistency to variable atmospheric escape from LHS 1140\,b, potentially driven by the changing XUV output of the host star.

He observations, though, are not the unique domain of ground-based high-resolution spectrographs --- detections of escaping He have been made with both the Hubble Space Telescope and JWST \citep[e.g.,][]{spake_helium_2018, mansfield_detection_2018, fu_water_2022, fournier-tondreau_near-infrared_2024, allart_complex_2025, ahrer_escaping_2025, krishnamurthy_continuous_2026}. In this work, we analyze four archival transits of LHS 1140\,b taken with the NIRISS instrument on JWST \citep{doyon_near_2023} to search for evidence of He escape from the planet. Although space-based instruments generally do not have the spectral resolution available to ground-based observatories, we will show below that the photometric stability of JWST NIRISS, in particular, can yield comparable limits on the presence of He to those from the ground. 

This work is structured as follows: we describe the observations and data analysis in Section~\ref{sec: Data Analysis}, and our search for He absorption signatures in Section~\ref{sec: He Modelling}. We then describe fits to the absolute stellar spectra in Section~\ref{sec: Stellar Analysis} to attempt to constrain possible variability timescales for LHS 1140. Finally, we provide a brief discussion and conclude in Section~\ref{sec: Discussion}.

% === Observations ====
\section{Observations \& Data Analysis} 
\label{sec: Data Analysis}

We downloaded all publicly available JWST observations of the LHS 1140 system taken with the Single Object Slitless Spectroscopy (SOSS; \citealp{albert_near_2023}) mode of the NIRISS instrument \citep{doyon_near_2023}. There were five such visits available at the time of writing: two transits of LHS 1140\,b from DD 6543 (PI: C.\ Cadieux --- previously published in \citealp{cadieux_transmission_2024}) taken on Dec 1, and Dec 25, 2023; two more transits from GO 7073 (PI: J.\ Lustig-Yaeger \& K.\ Stevenson) taken on Aug 10, 2025 and Jul 23, 2026; and one transit of LHS 1140\,c from Jul 26, 2025 also taken as part of GO 7073. We note that one further transit of planet b will be observed for GO 7073. 

The system has also been observed with other instruments on JWST (NIRSpec, MIRI), however, here we consider only the NIRISS observations as the other instrument modes do not have access to the metastable He triplet at 1.083\,µm. Hereafter, we will refer to the four transits of LHS 1140\,b as Visits 1--4 in chronological order.

We reduce each of the visits starting from the raw, uncalibrated data files using the \texttt{exoTEDRF} package \citep{radica_awesome_2023, feinstein_early_2023, radica_exotedrf_2024}. We closely follow the standard procedures laid out in e.g., \citet{radica_muted_2024, radica_promise_2025, radica_supersolar_2026}, including performing correction of 1/$f$ noise at the group-level, time-domain outlier flagging, and a piece-wise background removal. We flag and interpolate bad pixels using a spatial threshold of 10$\sigma$ and a temporal threshold of 5$\sigma$ from the surrounding median. Spatially, pixels are compared to a 3$\times$3 box, and temporally to the five neighbouring frames, and the $\sigma$ values above are the standard deviations of pixel values across these footprints. We then trace the target spectrum using the \texttt{edgetrigger} algorithm \citep{radica_applesoss_2022} and perform an aperture extraction using a width of 40 pixels. 

After extraction, we refine the wavelength solution for each visit by cross-correlating the median out-of-transit stellar spectrum with a NewEra stellar model \citep{hauschildt_newera_2025}, interpolated to match the properties of LHS 1140. This is to account for a potential offset in the wavelength solution of a given visit due to e.g., a sub-pixel shift in the position of the spectral trace along the detector's dispersion axis due to target acquisition imperfections. We use the \texttt{exoTHOMF} package \citep{radica_promise_2025, radica_exothomf_2026} to interpolate the model grid and assume the following parameters from \citet{cadieux_new_2024} for LHS 1140: $\rm T_{eff}=3096$\,K, $\rm \log g=5.041$\,cm/s$^2$, and $\rm [Fe/H]=-0.15$\,dex. We find slight sub-pixel shifts of 0.6, 0.2, and 0 pixels for Visits 2 through 4, respectively, and $-0.2$ pixels for the LHS 1140\,c visit. The spectrum was misplaced on the detector in Visit 1 due to erroneous target specification outside of the target acquisition field of view \citep{cadieux_transmission_2024}, so for this visit we re-use the bespoke wavelength solution derived by \citet{cadieux_transmission_2024}. We then perform absolute flux calibration following the methods of \citet{lim_atmospheric_2023, radica_promise_2025}.

% === 
\subsection{Light Curve Analysis}
\label{sec: Light Curve Analysis}

\begin{figure*}
    \centering
    \includegraphics[width=0.9\linewidth]{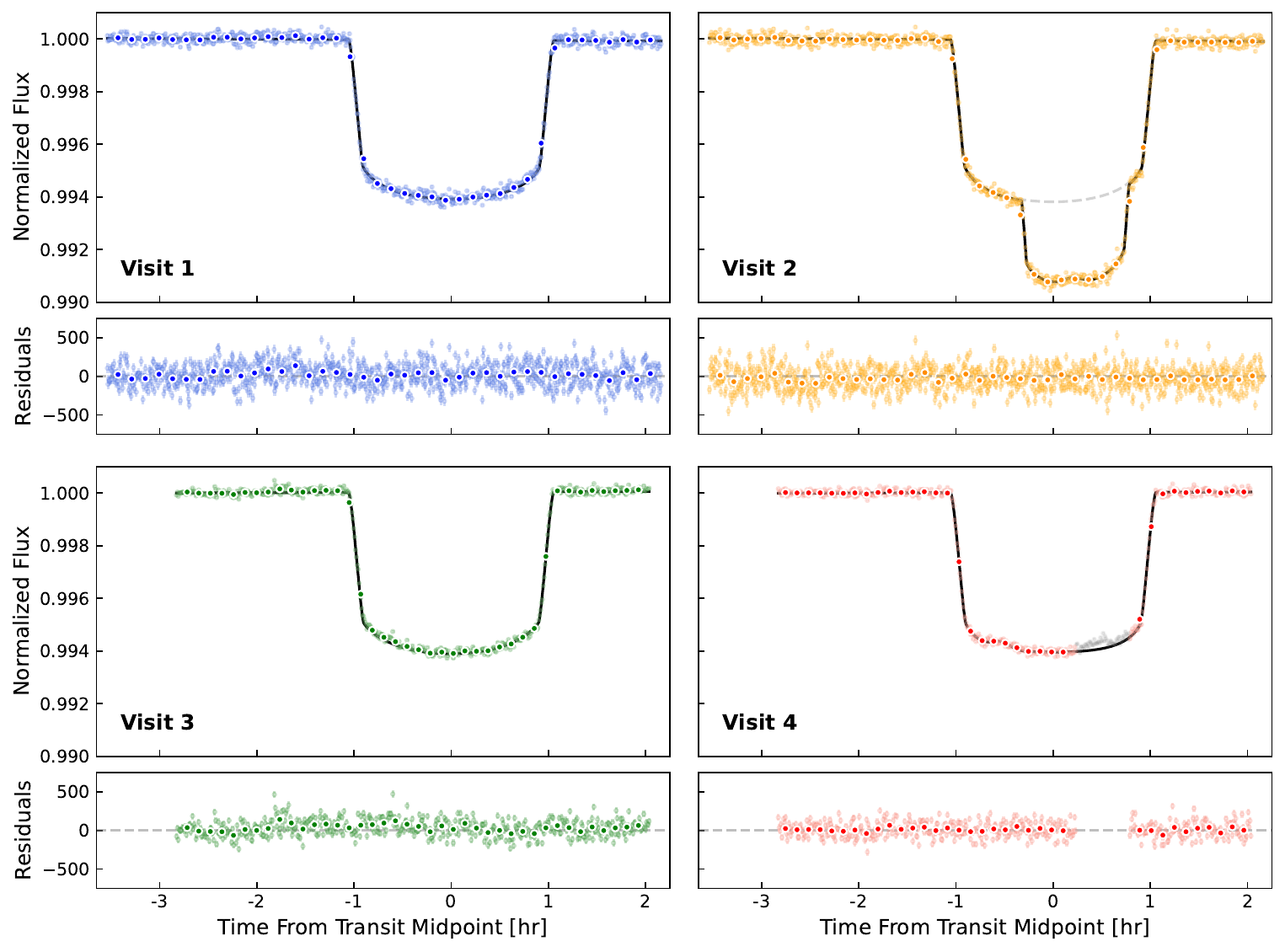}
    \caption{LHS 1140\,b NIRISS/SOSS white light curves for each of the four visits. White light curves are constructed using the first SOSS spectral order (0.85--2.85\,µm). Data are shown in coloured points and best-fitting transit models in black. The lower panels in each quadrant shows the residuals to the best fitting model. A serendipitous double-transit of planet c can be seen during the second visit. Grey points in Visit 4 show timestamps that were excluded from the fit due to the potential presence of a flare.}
    \label{fig: White Light Curves}
\end{figure*}

We construct white light curves for each of the four LHS 1140\,b transits by summing all flux in the first SOSS spectral order (0.85--2.85\,µm). We only consider the first spectral order as it has a significantly higher S/N than the second order (which spans 0.6--0.85\,µm) and contains the wavelengths corresponding to the metastable He triplet.

We use the flexible \texttt{exoUPRF} package \citep{radica_exouprf_2024} to fit a transit model to the white light curve of each visit separately. Our transit model consists of two parts: an astrophysical and systematics component. For the astrophysical model we use the \texttt{batman} package \citep{kreidberg_batman_2015} to calculate the transit shape. We fix the orbital period to 24.73723\,d \citep{cadieux_transmission_2024} and allow the mid-transit time, $T_0$, scaled semi-major axis, $a/R_s$, scaled planet radius, $R_p/R_*$, and orbital inclination, $inc$, to vary. We also fit the two parameters of the power2 limb-darkening law \citep{hestroffer_centre_1997, claret_power-2_2022} and assume a circular orbit. Priors and best-fitting values for each fitted astrophysical parameter can be found in Table~\ref{tab: WLC Parameters}, and agree well with those found by \citet{cadieux_transmission_2024}. For the systematics model, we fit a linear slope with time. We also include $f_e$, the normalized flux errors, as an additional free parameter in each light curve fit. We sample the posterior space using \texttt{dynesty} \citep{speagle_dynesty_2020} with 1000 live points. 

\begin{deluxetable}{cc|ccccc}
 \centering
 \tabletypesize{\scriptsize}
 \label{tab: WLC Parameters}
 \tablecaption{Best-fitting white light transit parameters}
 \tablehead{Parameter & Prior Range & Visit 1 & Visit 2 & Visit 3 & Visit 4 & Weighted Average \\
  & & Dec 1, 2023 & Dec 25, 2023 & Aug 10, 2025 & Jul 23, 2026 & 
  }
    \startdata
     Per [d] & Fixed & \multicolumn{5}{c}{24.73723} \\
     T0 [MJD] & $\mathcal{U}$[T0$\pm$2\,hr] & $60279.46094_{-0.00002}^{+0.00002}$ & $60304.19829_{-0.00002}^{+0.00002}$ & $60897.89220_{-0.00002}^{+0.00002}$ & $61244.21349_{-0.00002}^{+0.00002}$ & - \\
     $\rm R_p/R_s$ & $\mathcal{U}$[0.01, 0.9] & $0.0743_{-0.0002}^{+0.0002}$ & $0.0745_{-0.0001}^{+0.0001}$ & $0.0746_{-0.0002}^{+0.0002}$ & $0.0748_{-0.0001}^{+0.0001}$ & $0.0745_{-0.0001}^{+0.0001}$ \\
     $\rm a/R_s$ & $\mathcal{U}$[20, 100] & $94.60_{-1.13}^{+1.08}$ & $95.36_{-0.61}^{+0.51}$ & $93.75_{-1.06}^{+1.12}$ & $93.62_{-0.60}^{+0.85}$ & $94.40_{-0.38}^{+0.38}$ \\
     inc [deg] & $\mathcal{U}$[80, 90] & $89.89_{-0.04}^{+0.05}$ & $89.92_{-0.03}^{+0.03}$ & $89.85_{-0.03}^{+0.04}$ & $89.85_{-0.02}^{+0.03}$ & $89.87_{-0.01}^{+0.02}$ \\
     u1 & $\mathcal{U}$[$-$1, 1] & $0.23_{-0.03}^{+0.03}$ & $0.24_{-0.03}^{+0.03}$ & $0.29_{-0.03}^{+0.03}$ & $0.09_{-0.04}^{+0.05}$ & $0.23_{-0.02}^{+0.02}$ \\
     u2 & $\mathcal{U}$[$-$1, 1] & $0.14_{-0.05}^{+0.05}$ & $0.16_{-0.04}^{+0.04}$ & $0.03_{-0.05}^{+0.05}$ & $0.29_{-0.08}^{+0.08}$ & $0.13_{-0.03}^{+0.03}$ \\
     $f_e$ [ppm] & $\log \mathcal{U}$[$-$6, $-$1] & $142_{-4}^{+4}$ & $142_{-4}^{+4}$ & $140_{-3}^{+3}$ & $97_{-4}^{+4}$ & - \\
    \enddata
\end{deluxetable}

Visit 2 features a serendipitous transit of planet c, as well as a potential star spot crossing just after mid-transit. For this visit, we also include $T_0$, $a/R_s$, $R_p/R_*$, and $inc$ values for planet c and fix its period to 3.777940\,d \citep{cadieux_new_2024}. We model the spot crossing as a Gaussian with a freely fit amplitude, width, and position \citep[e.g.,][]{roy_diversity_2025, murphy_kronos_2026}. 

Visit 4 also features a spot crossing just after ingress which we model in the same manner as Visit 2. Furthermore, there is evidence for a potential small-amplitude flare event just after mid-transit. LHS 1140\,b is not known to be a regularly-flaring star \citep{medina_galactic_2022}. However, the structure of this event is quite different from that of the spot crossings seen in Visits 2 and 4 and similar to small-amplitude or micro-flare structures seen in JWST observations of other M dwarf stars \citep[e.g.,][]{piaulet-ghorayeb_jwstniriss_2024, ahrer_escaping_2025}. There is no clear structure in the H$\alpha$ light curves (contained within SOSS order 2) which is generally used as a concrete diagnostic of a flare \citep[e.g.,][]{lim_atmospheric_2023, radica_promise_2025, piaulet_strict_2025}, though the SOSS spectra are low-S/N at these wavelengths and partially contaminated by the spectral trace of a background star. Considering all of the above, we elect to be conservative and cut the integrations (no.\ 350 -- 405) corresponding to the potential flare.

In total we fit nine parameters to the white light curves for Visits 1 and 3, 16 for Visit 2, and 12 for Visit 4. The white light curves for each visit and the best fitting transit models are shown in Figure~\ref{fig: White Light Curves}.

\begin{figure*}
    \centering
    \includegraphics[width=0.9\linewidth]{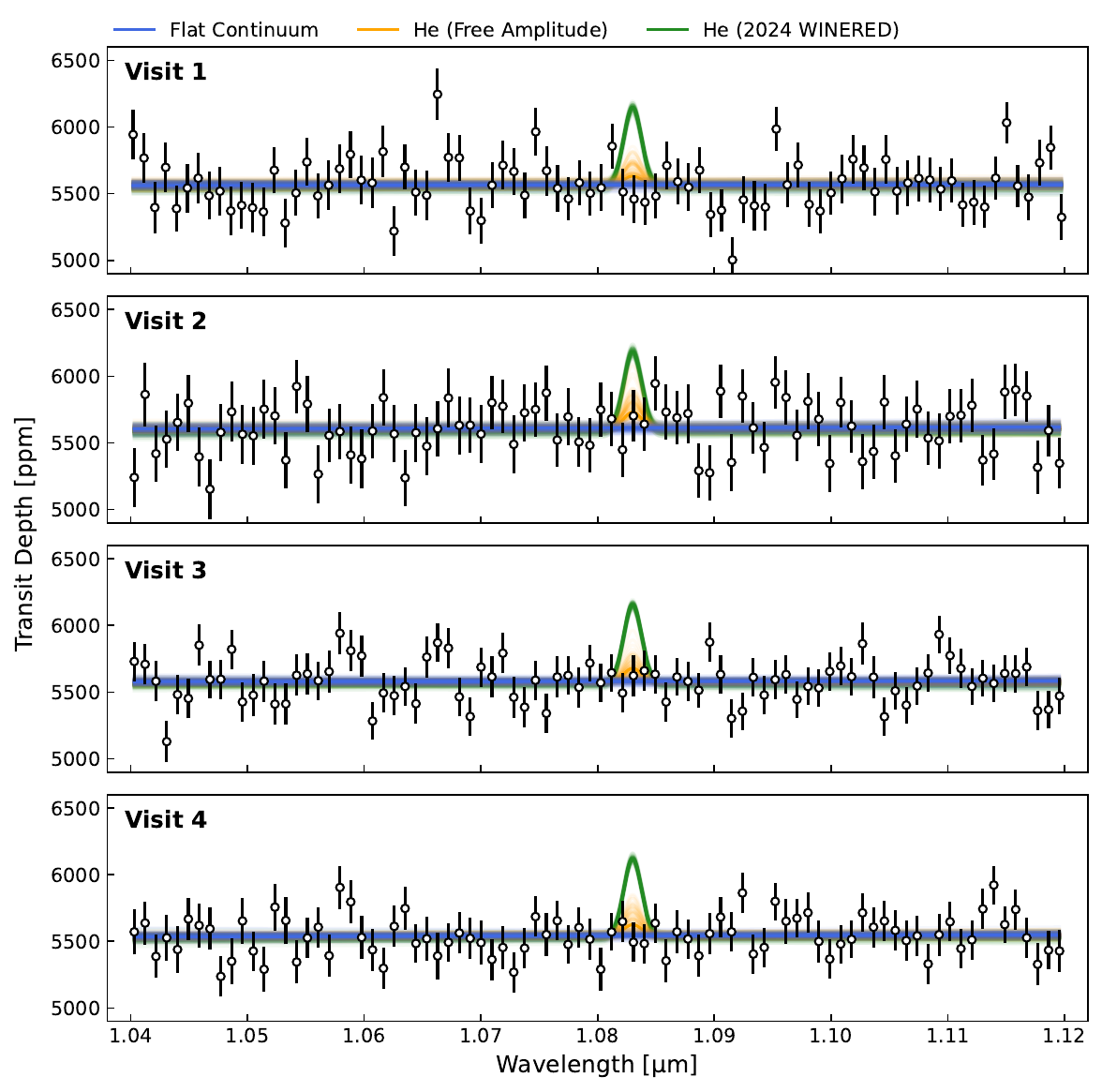}
    \caption{Transmission spectra of LHS 1140\,b for each of the four visits focused on wavelengths around the 1.083\,µm He triplet (which is unresolved at the spectral resolution of NIRISS/SOSS; $R$$\sim$600). Pixel-level transit depths (i.e., one transit depth per pixel column on the detector) are shown in black. Posterior draws for several model fits are shown with coloured lines: a flat continuum with no absorption lines (blue); a flat continuum with a Gaussian He absorption feature at 1.083\,µm (orange); a flat continuum with a Gaussian He absorption feature with its amplitude fixed to that observed by \citet{cherubim_helium_2026} using WINERED in 2024 (green).}
    \label{fig: He Spectra}
\end{figure*}

We then fit the spectrophotometric light curves at the pixel-level, that is, one light curve per pixel column on the detector. Since our analysis focuses on the 1.083\,µm metastable He triplet, we limit our fits to the 86 pixel columns spanning the wavelength range 1.04--1.12\,µm. We fix the scaled semi-major axis and orbital inclination to the weighted average of the four visits, and the mid-transit time to the best-fitting visit-specific value (Table~\ref{tab: WLC Parameters}). For Visit 2 we also fix the orbital parameters of planet c to their best-fitting values from the white light curve fit. We fit the two parameters of the power2 limb-darkening law using Gaussian priors centered on the predictions of \texttt{ExoTiC-LD} \citep{Grant2024ExoTiC-LD:Coefficients} with widths of 0.2 \citep{patel_empirical_2022}. We allow the scaled planet radius, systematics parameters, and normalized flux error terms to freely vary. We find normalized flux errors that decrease from $\sim$2200 to $\sim2000$\,ppm with increasing wavelength (following the instrument throughput) across the four visits. For visits with spot crossings, we fix the spot position and width to the best-fitting white light curve values and put a Gaussian prior on the amplitude, centered on the white light value with a width of 100\,ppm, to allow for wavelength evolution. In total, our spectrophotometric fits have six free parameters for Visits 1 and 3, eight for Visit 2, and seven for Visit 4. The final pixel-level transmission spectra for each visit are shown in Figure~\ref{fig: He Spectra}. We also test deriving the individual transmission spectra using a divide-white technique \citep[e.g.,][]{mansfield_detection_2018}, applying a quadratic trend to fit the out-of-transit baseline, and trimming the amount of baseline itself during the light curve fits but find virtually identical results. 

% === He Modelling ====
\section{Helium Line Modelling} 
\label{sec: He Modelling}

The primary goal of our analysis is to determine whether any of the four NIRISS/SOSS transits of LHS 1140\,b show evidence for He absorption. The He triplet is unresolved at the SOSS resolution ($R$$\sim$600) and will often only appear in one or two individual channels centered on 1.083\,µm \citep{fu_water_2022, ahrer_escaping_2025, allart_complex_2025}  As can be seen in Figure~\ref{fig: He Spectra}, the spectra appear quite flat in this narrow wavelength range and there is no clear He absorption signature for any visit. 

In order to quantify this observation, we undertake a Bayesian model comparison exercise. We fit three different models to each spectrum. The first is the null hypothesis which we model as a flat continuum with no spectral features. We also allow for the continuum to be sloped to account for the possibility of low-frequency spectral variation over this narrow wavelength range. For the second model we add a Gaussian He line on top of the continuum \citep[e.g.,][]{fu_water_2022, ahrer_escaping_2025, cherubim_helium_2026}. We fix the position of the Gaussian to 1.083\,µm and let its amplitude freely vary. We fix the full-width-half-max (FWHM) to 0.0018\,µm (18\,\AA), which is equivalent to the He line width of 0.84\,\AA\ observed by \citet{cherubim_helium_2026} in 2024 with WINERED when convolved down to the spectral resolution of NIRISS/SOSS. We note though, that since the He line is generally unresolved in NIRISS data, the assumed line width at high resolution has minimal impact on the final results. For the final model, we include a Gaussian He line on top of the continuum but fix both the line width and amplitude to that observed by \citet{cherubim_helium_2026} in 2024, again convolved down to the resolution of SOSS (amplitude=589\,ppm).

\begin{deluxetable}{cc|ccc}
 \centering
 \tabletypesize{\footnotesize}
 \label{tab: He Parameters}
 \tablecaption{Helium Line Fit Results}
 \tablehead{Visit & Model & He Amp & $\ln Z$ & $\ln B$ \\
  & & [ppm] & & (rel.\ to flat)
 }
    \startdata
     \multirow{3}{*}{1} & Flat Continuum & - & $-578.04$ & - \\
     & He (Free Amp) & $66.48^{+83.03}_{-48.55}$ & $-580.49$ & $-2.45$ \\
     & He (WINERED Amp)$^1$ & - & $-589.37$ & $-11.33$ \\
     \hline
     \multirow{3}{*}{2} & Flat Continuum & - & $-578.14$ & - \\
     & He (Free Amp) & $126.69^{+126.99}_{-89.05}$ & $-579.50$ & $-1.36$ \\
     & He (WINERED Amp)$^1$ & - & $-583.83$ & $-5.69$ \\
     \hline
     \multirow{3}{*}{3} & Flat Continuum & - & $-560.55$ & - \\
     & He (Free Amp) & $103.16^{+92.03}_{-70.87}$ & $-562.30$ & $-1.75$ \\
     & He (WINERED Amp)$^1$ & - & $-570.39$ & $-9.84$ \\
     \hline
     \multirow{3}{*}{4} & Flat Continuum & - & $-554.20$ & - \\
     & He (Free Amp) & $81.36^{+95.81}_{-57.70}$ & $-556.05$ & $-1.84$ \\
     & He (WINERED Amp)$^1$ & - & $-565.57$ & $-11.36$ \\
    \enddata
    \tablecomments{$^1$ Gaussian Helium amplitude fixed to 589\,ppm.}
\end{deluxetable}

We fit all three models to the transmission spectrum of each visit using the \texttt{dynesty} nested sampling routine \citep{speagle_dynesty_2020} using 1000 live points. Best-fitting models and random posterior draws are overplotted for each visit in Figure~\ref{fig: He Spectra}. The best-fitting He amplitudes (for the second model) as well as model comparison statistics are summarized in Table~\ref{tab: He Parameters}.

For all visits, the flat continuum model has the highest Bayesian evidence, indicating that it is the preferred model in each case. Quantitatively, a flat continuum lacking any He absorption is preferred over a model with He with a log-Bayes factor ranging from 1.36 to 2.45 (equivalent to odds ratios from 3.9:1 to 11.6:1 or $\sigma$$\sim$2.2--2.7 using the \citet{selke_calibration_2001} approximation). The model with the He amplitude fixed to the 2024 WINERED value is rejected much more strongly in favour of the flat continuum. Here, the log-Bayes factors range from 5.69 to 11.36 (equivalent to odds ratios from $\sim$300:1 to $\sim$8.6$\times$10$^4$:1 or $\sigma$$\sim$3.8--5.1 using the \citet{selke_calibration_2001} approximation). In summary, all four LHS 1140\,b transits rule out the presence of a He signal equivalent to that seen by \citet{cherubim_helium_2026} in 2024, and disfavour the presence of He absorption in general. 

\begin{figure}
    \centering
    \includegraphics[width=0.99\columnwidth]{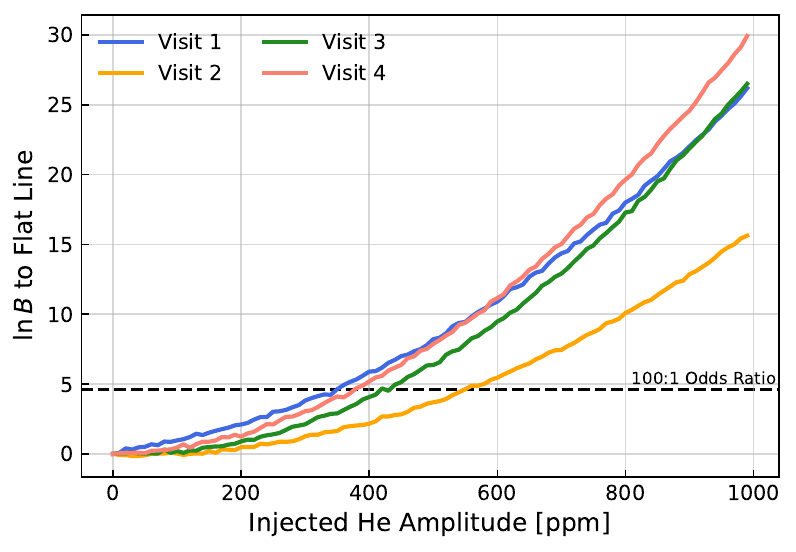}
    \caption{Results of He absorption injection-recovery tests for each of the four visits. Larger values of the Bayes factor ($B$) indicate stronger rejections of the He absorption model in favour of a flat continuum. }
    \label{fig: Inject Recover}
\end{figure}

It should be noted though that none of the JWST transits are contemporaneous with the WINERED detection in 2024, with the minimum time separation being $\sim$9\,months --- equivalent to the $\sim$1-year separation between the two WINERED visits presented by \citet{cherubim_helium_2026}. As noted by \citet{cherubim_helium_2026}, it is possible that the stellar XUV flux was higher during the epoch of the 2024 WINERED visit compared to both the epoch of the 2025 WINERED and the four NIRISS visits, resulting in a detectable atmospheric outflow only during this one epoch. Further monitoring of the system can confirm this hypothesis.  

To assess the limits that each transit can place on the presence of He absorption, we perform a series of injection-recovery tests. We fit a flat continuum + Gaussian He model to each transmission spectrum, keeping the He line width fixed to 18\,\AA\ as above, and altering the line amplitude from 0 to 1000\,ppm in steps of 10\,ppm. We then compare the Bayesian evidence value for each fit to the flat continuum case. These results are summarized in Figure~\ref{fig: Inject Recover}.

As expected, the degree to which He absorption is disfavoured increases monotonically with the amplitude of the injected He line. We place upper limits of 350, 550, 420, and 380\,ppm respectively for each of the four visits. We define our upper limits at a log-Bayes factor of 4.61 compared to the flat continuum, equivalent to an odds ratio of 100:1 or 3.5-$\sigma$ using the \citet{selke_calibration_2001} approximation. These correspond to He line amplitudes of 0.72\%, 1.15\%, 0.88\%, and 0.79\%, respectively, at the resolution of WINERED ($R$$\sim$68,000).

% === Stellar Spectrum ====
\section{Stellar Spectrum Analysis} 
\label{sec: Stellar Analysis}

\citet{cherubim_helium_2026} cite the detection of He escape from LHS 1140\,b in their 2024 WINERED dataset, and lack of a detection in 2025 as evidence for variable He escape, likely driven by a change in the host star's XUV flux. Since the five NIRISS visits (four LHS 1140\,b transits and one transit of planet c) to the system span nearly $\sim$2.5 years, we have high precision (though low resolution) spectra of LHS 1140 spanning a baseline of many stellar rotation periods (though with sparse time sampling). We thus attempted to fit the spectra of LHS 1140 itself to determine whether we could conclude any evidence for variability on the stellar surface (i.e., in the properties of the photosphere itself or of stellar surface heterogeneities), which could give insight into the evolution of the star's high-energy output. 

To this end, we again use \texttt{exoTHOMF} to compare our flux-calibrated stellar spectra to stellar atmosphere models. To create the stellar spectra, we take the median flux of all out-of-transit integrations, and use the standard deviation of the out-of-transit flux cube as the flux errors. Following e.g., \citet{moran_high_2023, radica_promise_2025, ahrer_escaping_2025}, we then create models of inhomogeneous stellar photospheres by interpolating over grids of NewEra stellar models. 

\begin{figure*}
    \centering
    \includegraphics[width=0.9\linewidth]{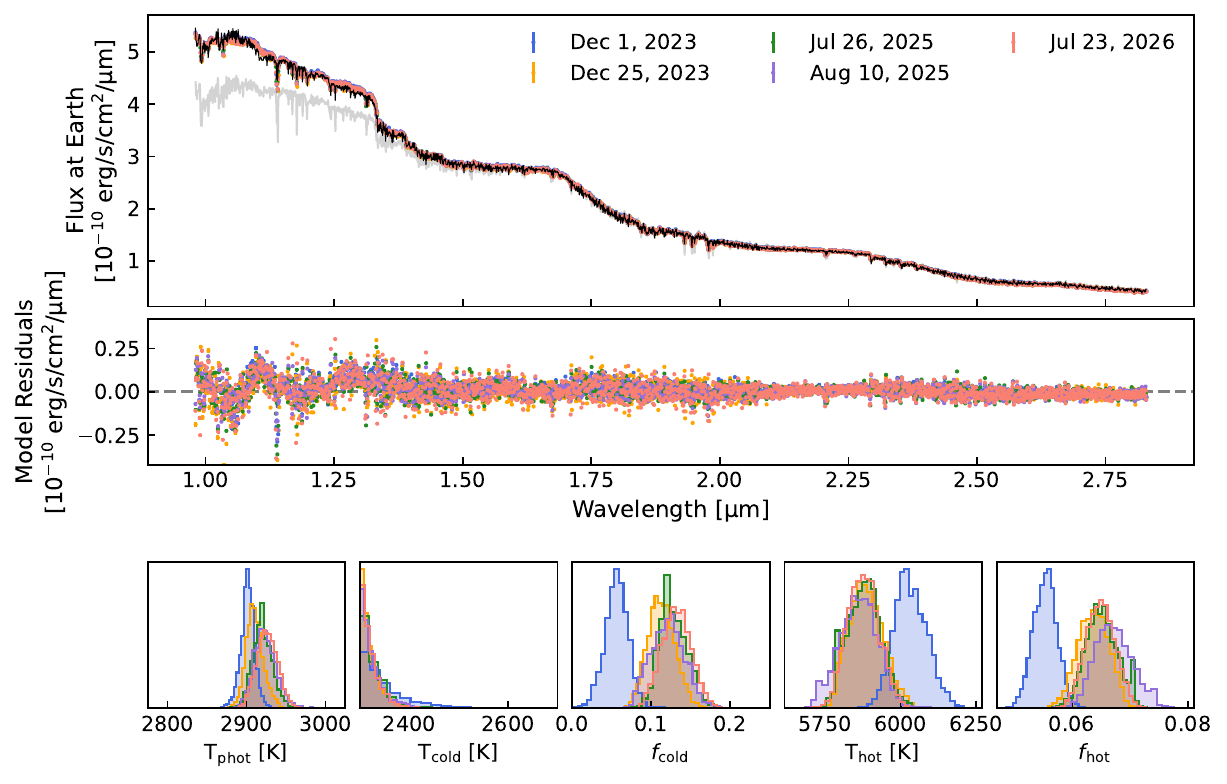}
    \caption{Results of stellar model fits to the out-of-transit stellar spectra. 
    \emph{Top}: Flux calibrated order 1 stellar spectra for the four LHS 1140\,b visits and one LHS 1140\,c are shown in coloured points and the best-fitting NewEra stellar model in black. In grey is a NewEra model using the stellar parameters from \citet{cadieux_new_2024}, which does not produce a satisfactory match to the data. 
    \emph{Middle}: Model residuals from each of the five visits. 
    \emph{Bottom}: Marginal posterior distributions for several key model parameters. The fits find consistent properties between the five visits, with evidence for both colder and warmer components in addition to the background photosphere. Note that 2300\,K is the coldest model available in the NewEra grid.}
    \label{fig: Stellar Spectra}
\end{figure*}

We allow for three distinct components in the models: a background photosphere with effective temperature $T\rm _{phot}$, and gravity, $\log g$; a cold component with temperature, $T_{\mathrm{cold}} < T\rm _{\mathrm{phot}}$ and covering fraction $f\rm _{cold}$; and a hot component with $T_{\mathrm{hot}} > T\rm _{\mathrm{phot}}$ and $f\rm _{hot}$. We further enforce the condition $f_\mathrm{hot} + f_\mathrm{cold} < 0.5$, and allow the hot and cold components to have gravities that can differ from the photospheric $\log g$ by at most 0.5. We put broad priors on all values with temperatures spanning 2300\,K (the lowest temperature for which NewEra models are available) to 7000\,K and $\log g$ from 3.0 to 5.5. The models are scaled by $R_*^2/D^2$, where $R_*=0.2159\,R_\odot$ is the stellar radius and $D=14.9861\,\rm pc$ is the distance to the system \citep{cadieux_new_2024}. We also include a flux scaling parameter in the fit (which can range from 0.9 to 1.1) to account for uncertainties in these values. Finally, we include an error inflation term that multiplies the flux errors. We sample the parameter space using \texttt{dynesty} \citep{speagle_dynesty_2020} with 1000 live points. We note that we only use data from the first SOSS order for the fits (wavelengths $>$0.85\,µm) as the second orders for most visits are contaminated by the spectra of background sources.

The results of these multi-component fits are summarized in Figure~\ref{fig: Stellar Spectra}. We also test fits with one- or two-component models (i.e., photosphere-only, or photosphere+cold/hot) and find that the three-component fits are strongly preferred in each case by $\ln B \gtrsim 20$ relative to the photosphere + hot or $\ln B \gtrsim 150$ relative to the photosphere + cold or photosphere-only fits. 

In general, we find a high degree of consistency in the fitted parameters across four of the five visits with their photosphere and heterogeneity components generally agreeing to within 1-$\sigma$ (Figure~\ref{fig: Stellar Spectra}). The fitted photospheric temperature is colder by about $\sim$200\,K than that found by previous studies \citep[e.g.,][]{lill-box_planetary_2020, cadieux_new_2024}, and we find evidence for a cold component at least 500\,K colder than the photosphere with a $\sim$10\% covering fraction, and a hot component $\sim$3000\,K warmer than the photosphere with a $\sim$7\% covering fraction. Visit 1 is generally more discrepant (though still consistent at the 2-$\sigma$ level with the other four visits). The first visit prefers a consistent temperature for its colder component, through a smaller covering fraction. Similarly it prefers a lower covering fraction for the hotter component, though a temperature that is $\sim$200\,K higher.

The inference of a cold component is limited by the low-temperature extent of the NewEra stellar model grid. To attempt to rectify this issue, we run fits using the SPHINX model grid \citep{iyer_sphinx_2023} which was designed to model M dwarf atmospheres and includes temperatures as low as 2000\,K. However, we were unable to achieve a satisfactory fit to the data using the SPHINX grid, in large part because the hottest model available in the grid is 4000\,K, and our data seem to require the addition of a spectral component at a significantly higher temperature. The inference of a high-temperature component is driven by wavelengths $<$1.5\,µm which have an excess of flux compared to a model at the published effective temperature of LHS 1140 (compare the data and grey model in Figure~\ref{fig: Stellar Spectra}).

\citet{cadieux_transmission_2024} also inferred the presence of a hot component on the photosphere of LHS 1140 via modelling the impacts of the TLS effect on their LHS 1140\,b transmission spectra. However, their modelling indicated a temperature $\sim$300\,K warmer than the photosphere, which is much cooler than what we find here.

It should be noted, though, that this genre of analysis comes with a variety of caveats, particularly when applied to late-type stars. Stellar models do a notoriously bad job of replicating the spectra of M dwarfs \citep[][Fig.~\ref{fig: Stellar Spectra}]{iyer_sphinx_2023, lim_atmospheric_2023, radica_promise_2025, piaulet_strict_2025, glidden_overestimated_2026}, spots and faculae have their own unique spectral features and cannot simply be modelled as a colder or hotter photosphere \citep[e.g.,][]{shapiro_curious_2026, murray_panchromatic_2026}. 

Nevertheless, though the absolute parameter values of our fits may be biased by the shortcomings described above, interpretation of the relative values should be on more stable footing. The fact that the inferred parameters for Visit 1 appear more discrepant with the other four visits (which are highly consistent) might be interpreted as evidence for variability. Further monitoring of the system is necessary to confirm this.

% === Discussion & Conclusions ====
\section{Discussion \& Conclusions} 
\label{sec: Discussion}

In this work, we analyzed four transits of LHS 1140\,b, spanning Dec 2023 to Jul 2026, to search for evidence for metastable He absorption. However, we do not find a He signature in any of the four transmission spectra, with each disfavouring the presence of He compared to a flat continuum by odds ratios ranging from 3.9--11.6:1. The data can also robustly rule out He absorption at the level seen by \citet{cherubim_helium_2026} in 2024 with odds ratios of 300--8.6$\times$10$^{4}$:1 compared to a flat continuum. Though, we again explicitly note that all of the JWST transits are separated by at least 9\,months from the WINERED detection.

We put strict limits on the amplitude of a He line that would have been observable at the epochs of the four transits (assuming it's width is comparable to that reported by \citealp{cherubim_helium_2026}). Via injection recovery tests we determine upper limits of 350, 500, 420, and 380\,ppm for Visits 1--4 respectively --- these are the amplitudes a He line would have to have at each epoch to be preferred by $\ln B=4.61$ (odds ratio of 100:1, comparable to 3.5\,$\sigma$ using the \citet{selke_calibration_2001} approximation) over a flat continuum. These values translate to 0.68\%, 1.07\%, 0.81\%, and 0.74\% at the spectral resolution of WINERED, and are comparable to the upper limit of 0.6\% derived by \citet{cherubim_helium_2026} from their 2025 WINERED spectrum. Such strict limits are possible as the extreme photometric stability of JWST observations compared to e.g., WINERED (which has to contend with observing through Earth's atmosphere) compensates for their lower spectral resolution. These results demonstrate that although high-resolution observations of He will always retain the edge due to their ability to fully resolve the line shape \citep{dos_santos_observing_2023}, space-based studies can achieve comparable detection sensitivities. 

We also fit the absolute stellar flux in each of the four LHS 1140\,b visits, as well as one additional spectrum from a NIRISS/SOSS transit of LHS 1140\,c to search for evidence of surface heterogeneities and variability which may give insights into the level and timescale on which the XUV environment of LHS 1140 changes. We find evidence for cold and hot components on the stellar photosphere, with four of the five visits having highly consistent properties. Visit 1 is significantly more discrepant, which could indicate variability, and further monitoring of the system is required to verify this.

In all, though we fail to detect any signatures of escaping He and thus cannot independently confirm the findings of \citet{cherubim_helium_2026}, our work provides a set of strict limits on potential He absorption which will be valuable to future long term analyses of the system's evolution and potential variable nature of atmospheric escape from LHS 1140\,b.

% ===== Extra Text ======
\begin{acknowledgments}
M.R.\ would like to acknowledge funding from the Natural Sciences and Engineering Research Council of Canada (NSERC), as well as the Canadian Space Agency (CSA). He would also like to thank J.\ Bean, C.\ Cherubim, A.\ Glidden, and C.\ Piaulet-Ghorayeb for helpful conversations, as well as K.\ Bennett, J.\ Lustig-Yaeger, and K.\ Stevenson for graciously agreeing to coordinate our publications.
This work is based on observations made with the NASA/ESA/CSA JWST.  The data were obtained from the Mikulski Archive for Space Telescopes at the Space Telescope Science Institute, which is operated by the Association of Universities for Research in Astronomy, Inc., under NASA contract NAS 5-03127 for JWST. The specific observations analyzed can be accessed via\,\dataset[10.17909/5nqn-4116]{10.17909/5nqn-4116}.
\end{acknowledgments}

% \begin{contribution}
% All authors contributed equally to this work.
% \end{contribution}

\vspace{5mm}
\facilities{JWST(NIRISS)}

\software{\texttt{astropy} \citep{astropy:2013, astropy:2018}, 
\texttt{batman} \citep{kreidberg_batman_2015},
\texttt{dynesty} \citep{speagle_dynesty_2020},
\texttt{exoTEDRF} \citep{radica_awesome_2023, feinstein_early_2023, radica_exotedrf_2024},
\texttt{exoTHOMF} \citep{radica_promise_2025, radica_exothomf_2026},
\texttt{exoUPRF} \citep{radica_exouprf_2024},
\texttt{ipython} \citep{PER-GRA:2007},
\texttt{jwst} \citep{bushouse_2023},
\texttt{matplotlib} \citep{Hunter:2007},
\texttt{numpy} \citep{harris2020array},
\texttt{scipy} \citep{2020SciPy-NMeth}
}

\bibliography{main.bib}{}
\bibliographystyle{aasjournalv7}

\end{document}